# Constrained RS coding for Low Peak to Average Power Ratio in FBMC - OQAM Systems

Job CHUNKATH, V.S. SHEEBA, Nisha VARGHESE

University of Calicut, India,
Department of Electronics & Communication Engineering,
Government Engineering College, Thrissur, India,
Corresponding author: e-mail: jobchunkath@gmail.com

*Abstract* – *Multi-carrier modulation techniques have now become a standard in many communication protocols. Filter bank based multi-carrier (FBMC) generation techniques have been discussed in the literature as a means for overcoming the shortcomings of IFFT/FFT based OFDM system. The Peak to Average Power Ratio (PAPR) is a problem faced by all multi-carrier techniques. This paper discusses the methods for reducing PAPR in a FBMC system while maintaining acceptable Bit Error Rate (BER). A new PAPR minimizing scheme called Constrained Reed Solomon (CRS) coding is proposed. The hybrid techniques using coding and companding are tested for different channel models and is found to yield promising results.*

## I. INTRODUCTION

The excessive growth of wireless communication services has resulted in an ever increasing demand for higher data rates in many application areas. The high data rate requirement has resulted in the exploration of new methods for implementing wireless communication. The present-day data rate requirement can be supported by the use of multi-carrier (MC) based systems, instead of single carrier (SC) systems prevalent in earlier days. The ability to perform well in multi-path propagation and frequency selective fading environments was considered as the major advantage of this concept. The multi-carrier communication scheme considers wideband frequency selective channel as a number of narrowband sub-channels with flat fading characteristics, which can be rectified by using simple equalization techniques at the receiver.

One of the popular multi-carrier systems that are presently being used is the Orthogonal Frequency Division Multiplexing (OFDM) technique [1]. Many wideband applications like Digital Audio (DAB), Digital Video Broadcast (DVB), and Wi-Fi make use of OFDM techniques. The main advantage of OFDM is its high spectral efficiency. It makes use of cyclic prefix (CP) to overcome the Inter-Symbol Interference (ISI) caused by delay spread. This redundancy in data results in decrease of overall data rate. The stop-band attenuation of OFDM system is -13dB, which causes leakage into adjacent sub-bands.

The above drawbacks of OFDM, has resulted in the search for a better system. This has resulted in the study of filter bank multi-carrier (FBMC) system as a possible alternative candidate for generation of orthogonal multi-carriers. In FBMC system it is possible to have prototype filters with large filter orders independent of the number of allotted sub-channels. The filters for different sub-channels are combined together with the help of a transmultiplexer [2]. Thus this system can have better stop-band attenuation which results in lower frequency leakage between adjacent sub-channels. The improvement in spectral shaping leads to the use of simpler equalization techniques at the receiver avoiding the use of CP. The FBMC scheme discussed in this paper makes use of Offset Quadrature Amplitude Modulation (OQAM) as the digital modulation technique for the data stream which is provided to the filter bank system. The OQAM modulation helps in maintaining orthogonality between the adjacent sub-channels, thus facilitating data recovery, even during adjacent channel interference. The system discussed in this paper makes use of OQAM digital modulation and filter bank based multi-carrier generation, hence can be called FBMC-OQAM system [3].

The FBMC system is also affected by PAPR problem. The conventional solution to this problem is amplifier '*back off* ' i.e., the amplifier operation characteristics are set below the desirable optimum values. Hence the amplifier will be able to operate in the linear region for a limited dynamic range of the signal, but this result in limited stage gain.

Diverse methods for PAPR reduction techniques for OFDM systems are described in the literature. The difference between OFDM system and FBMC system is that the overlapping adjacent symbols are present in an FBMC system [3]. Hence all the PAPR minimizing techniques used for OFDM are not applicable in an FBMC system. The Clipping and filtering method in conjunction with Tone Reservation (TR), Active Constellation Extension (ACE) and both these methods together (TRACE) are discussed in [4] by Eldessoki *et al*. The PAPR is reported at 5dB for the combined methods with BER of $10^{-6}$ for SNR greater than 14dB. The method discussed in [4] makes use of multiple stages to achieve PAPR reduction and the performance in fading channels is not discussed. Techniques like Multi-Block Joint Optimization (MBJO) [5] and sliding window tone reservation (SWTR) are used in FBMC-OQAM system for PAPR reduction in [6] by Shixian Lu *et al*. The BER performance of the systems is not reported, also the overlapping structure of FBMC-OQAM symbols lead to high system complexity. The clipping and its iterative compensation method is proposed for FBMC-OQAM system by Zsolt Kollar and Peter Horvath in [7], where complex receiver design for the compensation of clipping noise is the major drawback. PAPR reduction methods for an FBMC system with PAM symbols are discussed in [8], but this is limited only to PAM symbols. The *A*-law and $\mu$-law companding method is considered in [9], it is observed that BER performance is better when $\mu$-law companding is used. The PAPR reduction using DFT spreading and Identically Time Shifted Multicarrier (ITSM) conditioning for FBMC waveform is discussed by Dongjun Na and Kwonhue Choi in [10]. The PAPR value obtained is around 8dB. Junhui Zhao *et al* [11] discuss a joint optimization of Partial Transmit Sequence (PTS) and Improved Bi-Layer Partial Transmit Sequence with Iterative Clipping and Filtering (IBPTS-ICF) scheme for reducing PAPR to 4dB. A hybrid PAPR reduction scheme involving multi-data block Partial Transmit Sequence (PTS) and tone reservation (TR) is described in [12] by Han Wang *et al*. The lowest value of PAPR obtained is 6dB, but the BER performance in various channels is not discussed in [10], [11], and [12].

The reliability of a communication system depends on data integrity. The data integrity of a communication system is to be safeguarded by an efficient error correction scheme [13]. In this paper block coding techniques like Bose, Chaudhuri, and Hocquenghem (BCH) coding and Reed - Solomon (RS) coding are implemented. The RS code is sub-class of BCH code, it is a non-binary block code that can correct multiple random and as well as sporadic errors [14]. The non-binary nature of RS code can be effectively utilized for symbol transmission that is being used in multi-carrier communication. The RS coding scheme is finding increased popularity in mobile communication scenarios due to the efficient decoder implementation.

In this paper, we present an FBMC-OQAM system with different PAPR reduction methods. The multi-carrier signals from the transmitter are subjected to impairments of different channel models like ITU Vehicular A channel and Pedestrian B channel. Various PAPR reduction methods like non-linear companding are carried out along with error control coding techniques like BCH and RS coding. The different PAPR reduction methods are compared using CCDF plot. The system performance is evaluated using BER plot in different channel models.

The typical performance of FBMC-OQAM with constrained RS coding and $\mu$-law companding results in a PAPR of 4.6dB which is lower than the values claimed in recent papers [4], [7], and [10]. The proposed scheme confines PAPR to a narrow range of 0.55dB for a wide dynamic input channel load. Thus this method is helpful in preventing problems related to signal amplification due to non-linearity of final stage power amplifier. Thus the method proposed in this paper is capable of delivering a low PAPR and BER using the same coding scheme.

The paper is organized into five sections; section II, describes the system model. This section deals with the design and implementation of FBMC-OQAM system. The coding and companding technique for PAPR reduction are discussed in section III. The simulation results are presented in section IV. Section V summarizes the paper and its contributions.

## II. SYSTEM MODEL

The block schematic of the complete system is given in the Fig. 1. The transmitter section consists of input data block. This block applies necessary reformatting to the data as required by the error control coding scheme. The encoded stream of bits is then given to the FBMC-OQAM transmitter. The $\mu$-law companding is applied on the bit stream before transmission. At the receiver section, the first stage finds the inverse $\mu$-law transform. The subsequent FBMC-OQAM receiver is used to recover the data. The error correction on the recovered data is done using error control stage. The decoded data is then sent towards the data output block.

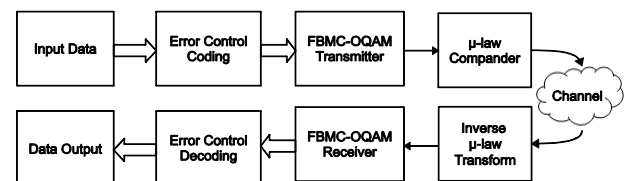

Fig.1 Block schematic of the proposed system.

### A. FBMC-OQAM System

One of the major concepts in the FBMC system is the use of transmultiplexer. The transmultiplexer is used for converting the signals from time division multiplexed (TDM) version into frequency division multiplexed (FDM) form and vice versa [15]. The FBMC- OQAM system is shown in Fig. 2. The main processing blocks in this direct form representation of an FBMC-OQAM system are OQAM pre-processing, synthesis filter bank, analysis filter bank, and OQAM post-processing [9].

### B. OQAM Pre/Post Processing

The first operation in OQAM pre-processing is the conversion from complex values to real values. The real and imaginary parts which are the constituents of the complex-valued QAM symbol $c_{k,l}$ where $k = 0,1,\ldots,M - 1$, are separated and time staggered by half the symbol period. The conversion from complex-to-real increases the sampling rate by a factor of 2.

The interference in adjacent sub-channels is avoided as the adjacent values in a single sub-channel and in the next sub-channel are multiplied by powers of $j$ (by $\theta_{k,n} = j^{(k+n)}$), hence they will be orthogonal to each other, thus ensuring interference-free transmission.

The first operation, in the OQAM post-processing is the multiplication of the sequence by $\theta^*_{k,n}$ (where $\theta^*$ is the conjugate of $\theta$), followed by the operation of separating the real part. The second operation is real-to-complex conversion, which decreases the sample rate by a factor of 2 [9].

### C. Synthesis & Analysis Filter Banks

All the sub-channel filters in synthesis filterbank $G_k(Z)$ with near perfect reconstruction (NPR) characteristics are formed from a single real valued linear phase FIR prototype filter $G_0(Z)$ with impulse response $p(m)$, by exponential modulation. The $k^{th}$ synthesis filter is defined by,

$$g_k(m) = p(m) exp\left(j\frac{2\pi k}{M}\left(m - \frac{L_p-1}{2}\right)\right) \quad (1)$$

where $m = 0, 1, \ldots, L_p - 1$ and $L_p$ is the filter length. The $k^{th}$ analysis filter can be realized by using the equation [9] which is defined by

$$f_k(m) = g^*_k(L_p - 1 - m)$$

In this paper, the prototype filter is designed by using frequency sampling method [9].

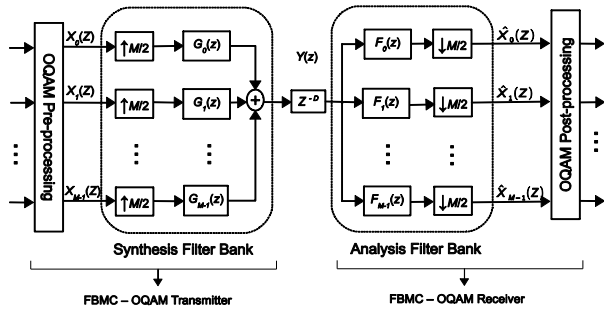

Fig.2 FBMC-OQAM system.

### III. CODING & COMPANDING

One of the techniques considered for PAPR reduction in the literature is block coding. The block coding results in spreading out the sequence of bits in such a way that the orthogonal signals will not have the same phase which prevents high signal peaks. This paper discusses the application of block coding followed by companding as an effective method for reducing PAPR. The block schematic of the system discussed is shown in the Fig. 1. The companding transform is helpful in curtailing high peak amplitudes in the modulated output signal, thus resulting in reduced PAPR. In this paper, the $\mu$-law companding method is implemented and it is found to deliver lower PAPR at acceptable levels of BER.

### A. $\mu$ − Law Companding

The output of $\mu$-law companding is given by the expression

$$F(y) = sgn(y)\frac{ln(1+\mu|y|)}{ln(1+\mu)}, \quad -1 \leq |y| \leq 1 \quad (2)$$

where $y$ is the input signal to the compander.

The companding transform is found to increase the dynamic range of signal, but it has a lesser effect on smaller amplitudes of the signals. The companding is carried out at the transmitter with a $\mu = 25$ [9].

At the receiver, the inverse of the companding transform is carried out on the received signal $r$. The expression of the inverse transform is given below.

$$F^{-1}(r) = sgn(r)\left(\frac{1}{\mu}\right)((1+\mu)^{|r|} - 1), \quad -1 \leq |r| \leq 1 \quad (3)$$

### B. Bose, Chaudhuri, and Hocquenghem Coding

The ability to detect and correct errors is an essential requirement for the successful performance of any communication system. The BCH codes are error correcting codes used to correct random errors occurring during data transmission. The BCH codes are linear, cyclic codes capable of correcting several errors in a block of message bits. Typically for any positive integer $m \geq 3$ and hamming distance $d_{min} \geq 2t + 1$, a binary BCH code having following parameters is found to exist.

Block length: $n = 2^m - 1$
Number of parity check digits: $r = n - k \leq mt$
Minimum distance: $d_{min} \geq 2t + 1$

where $\boldsymbol{n}$ is the length of encoded message bits, $\boldsymbol{k}$ is the input vector length, and $\boldsymbol{t}$ is the number of errors that can be corrected by the BCH code. The BCH encoding results in the introduction of $\boldsymbol{n - k}$ parity bits in the encoded message. This sequence of bits increases the Hamming distance between successive message bits. The increased distance and orthogonality of encoded message prevents the accumulation of the same phase and frequency signals which reduces the PAPR. The BER performance of the BCH coding is also good, due to the inherent error correcting capability. In this paper a BCH code with the parameters $(\boldsymbol{n, k}) = (\boldsymbol{127, 85})$ is chosen.

### C. Reed-Solomon (RS) Coding

The Reed - Solomon (RS) code is a block code scheme that utilizes the group of bits or symbols instead of bits as in the case of BCH code. If $\boldsymbol{k}$ symbols of a message are to be encoded with $\boldsymbol{r}$ parity symbols, it will form a codeword of length $\boldsymbol{n = k + r}$. As RS code utilizes symbol grouping and the number of symbols in a codeword is fixed as $\boldsymbol{n = 2^m - 1}$. The symbol error correcting capability of the code is $\boldsymbol{t = r/2}$. In this paper we also evaluate the suitability of RS (25, 16), which is a punctured form of RS (31, 19), for PAPR reduction and error correction.

### D. Constrained Reed-Solomon (CRS) Coding

The RS coding scheme suitable for FBMC system can be developed as follows, consider a $M$ channel FBMC system, where $M = 2^N$. For an OQAM scheme 2 bit symbols can be applied to a channel, hence total number of bits that can be sent using $M$ channels is $2^{N+1}$ bits.

Assuming that an error control coding scheme having a total size $\approx 2^{N+1}$ bits exists i.e. considering both message and parity bits together. Let $R_b$ be the number of parity bits of this coding scheme such that $R_b \leq 2^N$.

Consider a parity symbol is formed by $q$ bits from the available parity bits $R_b$. A maximum of $r$ parity symbols can be formed with $q$ bits each if

$$r \times q \leq 2^N \quad (4)$$

The total number of possible message bits, $K_b$ is

$$K_b = 2^{N+1} - r \times q \quad (5)$$

Assuming that each message symbol is formed by $p$ bits from the available $K_b$ message bits. A maximum of $k'$ message symbols can be formed with $p$ bits each, as

$$k' \times p \leq 2^{N+1} - r \times q \quad (6)$$
$$\text{i.e.} \quad k' \times p + r \times q \leq 2^{N+1} \quad (7)$$

Now consider a $(n, k)$ RS code, with a total symbol size

$$n = k + r \quad (8)$$

A constraint $p < q$, is applied on equation (7), keeping both $p$ and $q$ as integers. Then the upper limit for $p$ can be found from the relation

$$p \leq \frac{2^{N+1} - r \times q}{k'} \quad \text{where } k' \leq k \quad (9)$$

The lower limit of $p$ can be found by applying the condition $k' = k$ in the equation (9).

Applying the constraint $p < q$ on a RS (31, 19) code with $N = 6, r = 12, q = 5, k' = 16$ and evaluating equation (9), The value of $p$ is in the interval $3.58 < p \leq 4.25$, hence an integer value of $p = 4$ can be accepted.

The application of $p < q$ and $k' > r$ in the proposed constrained RS coding results in

$$k' \times p \cong r \times q \quad (10)$$

The conventional RS coding cannot be directly applied to the above FBMC system, hence a punctured and shortened RS (25, 16) is found suitable. This can be analyzed by substituting the values $k = 16, r = 9, p = q = 5$, which results in

$$k \times p \gg r \times q \quad \because k \gg r \text{ and } p = q \quad (11)$$

Thus it is evident that the proposed constrained RS coding has better message to parity bit ratio than conventional RS coding and hence better error correction capability.

The proposed Constrained Reed-Solomon code is found to result in the minimization of the total number of bits used when compared to conventional RS code. The different possible coding schemes with $n = 31, k \geq 19$ are explored in this paper. As $k$ is varied from 19 to 29, it is observed that the PAPR at the output of CRS encoder is found to decrease. This is due to the fact that 16 message symbols are distributed over an increased symbol space, which results in lower information content and subsequently lower PAPR. A comparison of PAPR obtained at the output of conventional RS encoder and CRS encoder when $k$ is varied from 19 to 29 is given in Table I. The PAPR obtained by using CRS code shows a steady decline with increasing value of $k$ even under full load conditions.

TABLE I
Comparison of PAPR and information at the encoder Output for $19 \leq k \leq 29$ under full load condition

| (n,k) | RS code PAPR (dB) | CRS code PAPR (dB) | Information ($I$) at the output of CRS encoder |
|---|---|---|---|
| (31,19) | 18.91 | 17.03 | 61 |
| (31,21) | 18.91 | 16.63 | 59 |
| (31,23) | 18.91 | 16.43 | 57 |
| (31,25) | 18.91 | 15.95 | 55 |
| (31,27) | 18.91 | 15.83 | 53 |
| (31,29) | 18.91 | 15.63 | 51 |

*E. Codeword Separation*

The separation between valid codeword is an important aspect that determines error correction capability. For BCH (127, 85), the probability of a valid codeword among the total bit combinations is $2^{-42}$. The encoded bits are transmitted as 128 bits, hence the probability of the valid codeword in the transmitted bits is $2^{-43}$. The Reed-Solomon codes belong to a class of linear block codes called the Maximum Distance Separable (MDS) codes. This results in the maximum error detection and correction capacity for the given $(n, k)$ combination. In conventional RS (31, 19) coding, the probability of a valid codeword among the total bit combinations is $2^{-60}$. The total valid codewords in a CRS scheme is $2^{64}$, and the number of bits transmitted is 128, hence the probability of a valid codeword among possible bit combinations in the CRS scheme is $p_{CRS} = 2^{-64}$. Thus it can be shown that codeword separation of the CRS scheme is the largest when compared to BCH and RS codes.

$$p_{CRS} < p_{RS} < p_{BCH} \quad (12)$$

For an encoding system, the probability at which a valid message occurs at the output can be related to the information $I_l$, and can be measured. The information contained in the $l^{th}$ message with probability $p_l$ is denoted as $I_l$, and defined as in equation 5.

$$I_l = \log_2\left(\frac{1}{p_l}\right) \quad (13)$$

The variation of PAPR with information for different values of $k$ is given in Table I. From the table it can be seen that CRS $(31, k)$ coding can minimize the PAPR of an encoded message. It can also be inferred from the values that the PAPR of an encoder output is related to the measure of information present

IV. SIMULATION RESULTS

*A. Simulation Setup*

The communication system is simulated with a transmitter /receiver and different channel models. The simulation is realized by using MATLAB®. A uniform distribution of $10^6$ random binary bits are generated and used as test data. The simulation is carried out on different ITU channel models. The FBMC-OQAM system with 64 sub-channels is tested on all three channel models and baseline performance is verified. The PAPR reduction methods like $\mu$-law companding, BCH and CRS coding jointly with $\mu$-law companding are tested. The system performance is evaluated by plotting the Bit Error Rate (BER) versus Signal to Noise Ratio (SNR) curve for different schemes. In this paper performance under six different methodologies are compared.

*B. Peak to Average Power Ratio*

The block coding methods investigated in this paper are BCH, punctured RS and the proposed CRS schemes. The extent of error correction and its impact on PAPR is also investigated for all these methods. In this paper a 64 sub-channel system is considered.

One of the error correction method used in this paper is the BCH (127, 85) coding. The message when encoded in this method produces 127 output bits which can be suffixed with one bit, so that 128 bits can be made available for OQAM modulation. It can be observed from Fig. 3 that the PAPR for random load is 4.2dB for BCH (127, 85) with $\mu$-law companding method. The Fig. 4 shows the PAPR at full channel load, under this condition the BCH coding scheme has PAPR of 20.41dB. Thus for a varying channel load the output PAPR widely varies for BCH coding scheme, hence Reed- Solomon coding scheme is explored.

The conventional RS (31, $k$) coding cannot be adapted directly to the problem as a 64 sub-channel system is to be realized; hence a punctured and shortened RS (25,16) code is implemented in this paper. This coding scheme with $\mu$-law companding offers a PAPR value of 4.2dB for random load and 5.99dB at full load, which is a variation of 1.79dB only.

The capability of CRS (31, $k$) code for reducing PAPR is also investigated. The number of sub-channels utilized in this paper is met by CRS (31, $k$) coding scheme with $k \geq 19$. The CRS (31, 19) code is used in further studies due to its better error correction capability. To investigate the performance of different schemes discussed, the PAPR is also obtained with all channels fully loaded and the values obtained under these conditions are given in Table II.

TABLE II
Typical valuesof $PAPR_0$ for various channel loads

| Methodology | Random load $P_0$ (dB) | Full load $P_0$ (dB) | Range (dB) |
|---|---|---|---|
| FBMC-OQAM | 14.4 | 18.92 | 4.52 |
| BCH (127,85) | 10.53 | 20.41 | 9.88 |
| RS (25,16) | 10.24 | 15.97 | 5.73 |
| CRS (31,19) | 11.74 | 12.76 | 1.02 |
| $\mu$-law companding | 6.78 | 18.92 | 12.14 |
| BCH + $\mu$-law companding | 4.2 | 20.41 | 16.21 |
| RS (25,16) + $\mu$-law companding | 4.2 | 5.99 | 1.79 |
| CRS (31,19) + $\mu$-law companding | 4.6 | 5.15 | 0.55 |

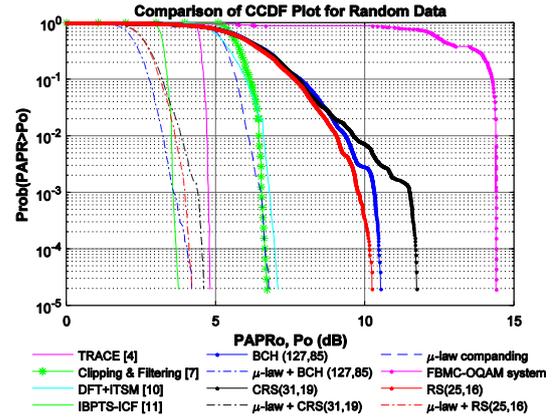

Fig. 3 Comparison of PAPR obtained by different techniques for random bit stream.

In the Fig. 3, the PAPR obtained by different methods are compared with the values claimed in recent papers [4], [7], and [10]. It is observed that PAPR reduction obtained by IBPTS-ICF [11] method is the only method that has lower value than the methods proposed in this paper. The Improved Bi-Layer Partial Transmit Sequence and Iterative Clipping and Filtering (IBPTS-ICF) scheme presented in [11] is a complicated approach when compared to the proposed CRS coding scheme which reduces the PAPR and also does error correction.

The PAPR values obtained at full load condition can be considered as the upper bound of the above system, for random load condition this value may be lower as shown in column two of Table II. The PAPR is observed to decrease when CRS coding and CRS coding with $\mu$-law companding are applied. The PAPR obtained under full load and random load conditions are 5.15dB and 4.6dB respectively. Thus the proposed CRS (31, 19) coding scheme has the least variation in PAPR irrespective of dynamic input channel load. This narrow range bound variation in PAPR prevents the RF power amplifier non-linearity from affecting the signal integrity.

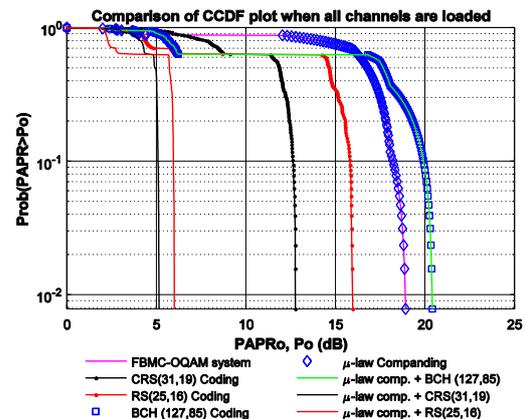

Fig. 4 Comparison of PAPR obtained by different techniques at full channel load.

*C. Bit Error Rate*

The BER performance of the system for various ITU channels is as shown in Fig. 5 and 6. It can be observed from Fig. 5 and 6 that CRS (31, 19) coding with $\mu$-law companding gives lower BER at lower SNR values for Pedestrian B and Vehicular A channel model. It is evident that the BER performance is the best when CRS (31, 19) coding is applied to FBMC-OQAM system. Thus, it can be inferred that the proposed CRS (31, 19) code performs better than conventional punctured RS (25, 16) and BCH (127, 85) codes.

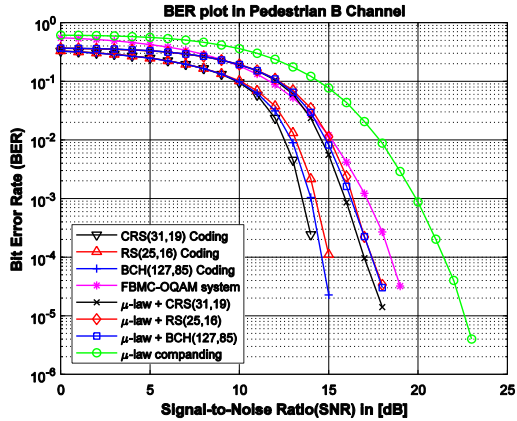

Fig. 5 Plot of BER vs. SNR obtained by different schemes for Pedestrian B channel model.

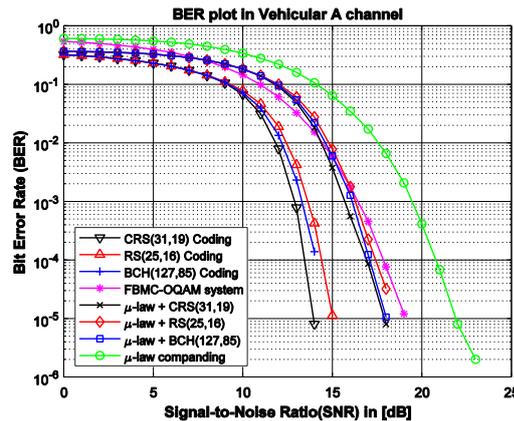

Fig. 6 Plot of BER vs. SNR obtained by different schemes for Vehicular A channel model.

V.   CONCLUSION

The two important parameters of a multi-carrier communication system are its BER and PAPR. This paper investigated the joint application of block coding and companding as a means of decreasing BER and PAPR. The different results obtained by using various methods were compared and it is found that BER performance of CRS scheme is found to be better for Pedestrian B and Vehicular A channel models.

The proposed constrained RS coding scheme discussed in this paper has the following results,
- The forward error correction and PAPR minimization is achieved together using the same constrained RS coding scheme.
- The worst case PAPR values of different coding schemes are compared for the first time.
- The PAPR of a system is proportional to the amount of information present at the encoder output.
- The proposed method has resulted in a PAPR of 4.6dB for FBMC-OQAM system, which is lower than the values claimed in recent papers [4], [7], and [10].
- The constrained RS coding method has confined the PAPR to a narrow range of 0.55dB for a wide dynamic input channel load, thus preventing the power amplifier non-linearities from affecting the signal amplification.

Thus the method presented in this paper is suitable for enhancing the performance of multi-carrier systems.

REFERENCES

[1] Abraham Peled and Antonio Ruiz, "Frequency domain data transmission using reduced computational complexity algorithms", IEEE international conference on Accoustics and Speech, Vol. 5, 1980, Page(s): 964-967.
[2] M. Vitterli, "Theory of multirate filter banks", IEEE Transactions on Accoustics, Speech and signal processing, Vol. 35, No. 3, March 1987, Page(s): 356-372.
[3] Pierre Siohan, "Analysis and Design of OFDM/OQAM systems based on filterbank theory", IEEE transactions on Signal Processing, Vol. 50, 2002, Page(s): 1170-1183
[4] S. Eldessoki, J. Dommel, K. Hassan, L. Thiele and R. F. H. Fischer, "Peak-to-Average-Power Reduction for FBMC-based Systems", WSA 2016; 20th International ITG Workshop on Smart Antennas, Munich, Germany, 2016, pp. 1-6.
[5] Daiming Qu, Shixian Lu and Tao Jiang, "Multi-Block Joint Optimization for the Peak-to-Average Power Ratio Reduction of FBMC-OQAM Signals", IEEE Transactions on Signal Processing, Vol. 61, No. 7, April 1, 2013, Page(s): 1605-1613.
[6] Shixian Lu, Daiming Qu and Yejun He, "Sliding Window Tone Reservation Technique for the Peak-to-Average Power Ratio Reduction of FBMC-OQAM Signals", IEEE Wireless Communications Letters, Vol. 1, No. 4, August 2012, Page(s):268-271.
[7] Zsolt Kollar and Peter Horvath, "PAPR Reduction of FBMC by Clipping and Its Iterative Compensation", Journal of Computer Networks and Communications, Volume 2012.
[8] Mishmy T. S. and Sheeba V. S., "Peak to Average Power Ratio reduction in Filterbank Based Multicarrier system", IEEE International conference on Advances in Computing and Communications, August 2013, Page(s): 377-381.
[9] Varghese,N; Chunkath,J; Sheeba, V.S, "Peak-to-Average Power Ratio Reduction in FBMC-OQAM System", Advances in Computing and Communications (ICACC), 2014 Fourth International Conference on , vol., no., pp.286,290, 27-29 Aug. 2014.
[10] Dongjun Na and Kwonhue Choi, "Low PAPR FBMC ", IEEE Transactions on Wireless Communications, Vol. 17, No. 1, pp 182-193, January 2018.
[11] Junhui Zhao, Shanjin Ni And Yi Gong, "Peak-to-Average Power Ratio Reduction of FBMC/OQAM Signal Using a Joint Optimization Scheme", Special Section on Physical and Medium Access Control Layer Advances in 5G Wireless Networks, IEEE Access, Volume 5, pp. 15810-15819,May 2017.
[12] Han Wang, Xianpeng Wang, Lingwei Xu and Wencai Du, "Hybrid PAPR Reduction Scheme for FBMC/OQAM Systems Based on Multi Data Block PTS and TR Methods", Special Section On Green Communications And Networking For 5G Wireless, IEEE Access, Volume 4, pp. 4761-4768, September 2016.
[13] Vijay K. Bhargava; Qing Yang; David J. Peterson,"Coding Theory and its Applications in Communication Systems", Defence Science Journal, Vol 43, No 1, January 1993, pp 59-69.
[14] Sklar, B., "Channel Coding: Part 3," in Digital Communications: Fundamentals and Applications, 2nded, Pearson Education India, 2009, pp 438-455.
[15] P. P. Vaidyanathan, "Fundamentals of Multirate Systems," in Multirate Systems and Filter Banks, Pearson Education India, 2006, pp 100-163.